\documentclass[aps,prl,twocolumn,10 pt,superscriptaddress,showpacs]{revtex4}
\usepackage{amssymb}

\usepackage{graphicx}

\begin{document}

\title{Voltage noise and surface current fluctuations in the superconducting surface sheath}
\author{J. Scola}\author{A. Pautrat}\author{C. Goupil}
\affiliation{CRISMAT/ENSI-Caen, UMR 6508 du CNRS,6 Bd Marechal
Juin, 14050 Caen, France.}
\author{L. M$\acute{e}$chin}
\affiliation{GREYC/ENSI-Caen, UMR 6072 du CNRS, 6 Bd Marechal
Juin, 14050 Caen, France.}
\author{V. Hardy}\author{Ch. Simon}
\affiliation{CRISMAT/ENSI-Caen, UMR 6508 du CNRS,6 Bd Marechal
Juin, 14050 Caen, France.}

\begin{abstract}
We report the first measurements of the voltage noise in the
surface superconductivity state of a type-II superconductor. We
present strong evidences that surface vortices generates surface
current fluctuations whose magnitude can be modified by the
pinning ability of the surface. Simple two-stage mechanism
governed by current conservation appears to describe the data. We
conclude that large voltage fluctuations induced by surface
vortices exist while the bulk is metallic. Furthermore, this
experiment shows that sole surface current fluctuations can
account for the noise observed even in the presence of vortices in
the bulk.
\end{abstract}


\newpage
\maketitle

 Separating bulk and surface effects is an ubiquitous problem in condensed matter physics.
For example, Flicker noise in semi-conductors \cite{vanderziel}
and $1/f$ like-noise in metals \cite{dutta,zimmerman} can occur
from either bulk or surface localized sources. Similarly, one can
cite the long standing debate about surface versus bulk effects in
driven non-linear systems like charge density waves \cite{gruner}
or vortex lattice in superconductors \cite{zeldov,nono}. Such
debate extends to the origin of the fluctuations responsible for
their electronic noise. In any case, the key question is how to
determine the relevant sources of disorder. From a technological
point of view, noise is a limiting factor for most of
applications, and it is necessary to know its origin before
expecting its minimization. It is also of theoretical interest to
know if, when analyzing disordered systems, boundaries effects are
of first importance or if they can be neglected compared to the
pure bulk treatment of the problem. A similar question has lead to
the proposition that bulk impurities play no major role in the
broad band noise generation of charge density waves \cite{ong}. As
a model system, the case of superconducting vortex lattices can be
particularly instructing.

The dynamics of a bulk vortex lattice has been heavily studied in
the literature. It is both understood and experimentally shown
that the vortices, when submitted to an overcritical current $I >
I_c$ and under steady state conditions, move in the bulk of the
sample with a well defined average periodicity \cite{echo}. This
motion is associated with a resistance $R_{ff} \leq R_n$ and an
electrostatic field $E = -V_L \wedge \omega$ ($V_L$ is the line
velocity and $\omega$ the vortex field) \cite{jos}.
Experimentally, it was also shown that $E$ can be separated into
its mean $\langle E \rangle$ and its fluctuating part $\delta
E^{*}$ \cite{clem}. The latter is the electric field noise.
Combined with the V-I relation $V = R_{ff} (I-I_{c})$, one can
easily realize that this noise can be expressed as fluctuations in
the bulk current $(I-I_c)$, or as fluctuations in the resistance
$R_{ff}$. Most of the theories invoke bulk pinning centers through
their interaction with the moving vortices, leading to vortex
density fluctuations and to their associated  $ \delta R_{ff}^{*}$
fluctuations \cite{clem}.
 Alternatively, noise may arise from over-critical current fluctuations $\delta (I-I_c)^{*}$, while $R_{ff}$ is constant \cite{placais}. For the simple reason of current conservation, $\delta (I-I_c)^{*}=\delta I_c^{*}$. The fluctuations
 are thus linked to the underlying pinning mechanism.
Cross-correlation magnetic flux noise experiments and rigorous
analysis of the correlation terms are in favor of a surface
origin, at least in Pb-In samples \cite{placais}. If fluctuations
are pure surface currents, it should be realized that the bulk is
only a host which transports the (here noisy) information.

As a consequence, the study of pure surface current fluctuations
coexisting with a "noise-free" (metallic) bulk should bring strong
experimental evidences on the validity or irrelevancy of this
scenario. In the surface superconductivity state, surface currents
naturally coexist with a metallic bulk. This means that the
$\delta R^*_{ff}$ term should be negligible above $B_{c2}$, so
that only current fluctuations can account for the possible
voltage noise. To our knowledge, the noise in the surface
superconductivity state has not been reported so far. We show here
that large voltage fluctuations do exist in this regime. Since it
is made clear that they originate from the surface (approximately
0.01\% of the total volume), a demonstration of the relevance of
the two-stage surface/bulk noise mechanism is brought. These
fluctuations exhibit the same magnitude as it is observed in the
conventional vortex state of type II superconductors and can be
modeled by similar arguments of statistical averaging.

 The prediction of Saint James and De Gennes \cite{degennes} that
superconductivity persists in a surface sheath up to a field
$Bc_{3} \approx 1.69 Bc_{2}$ has been confirmed by many
experiments. In particular, surface superconductivity is very
frequently observed in low kappa superconducting metals.
Consequently, a strip of pure Niobium ($T_c = 9.17 K$,
Ginzburg-Landau parameter $\kappa \approx 1$, coherence length
$\xi \approx 30$ nm) \cite{desorbo} has been chosen for this
experiment. For all measurements, the magnetic field $\mu_0 B$ was
applied perpendicular to the large faces. Note that, strictly
speaking, superconductivity nucleates on the typical scale of
$\xi$ over surfaces parallel to $\mu_0 B$. In practice, this
condition is locally realized by non-zero normal component thanks
to usual surface roughness, explaining the possibility of
important surface superconducting currents close to the largest
surfaces. Kulik calculated that this surface sheath is populated
by quantized flux spots ("Kulik's vortices" \cite{kulik}). It is
very likely that the local surface disorder acts as the pinning
potential for this short vortices (\cite{patrice} and references
herein), leading to a surface critical state. Therefore, the
associated superficial critical current $i_c (A/m)$ comes from a
pinning mechanism, and is consequently observed lower than its
theoretical upper bound \cite{hart}.  For $I \geq I_c\approx
2W.i_c$ (W=width), surface vortex depinning occurs. Any excess
current ($I-I_{c}$) is transported in the bulk by normal
electrons. It is important to realize that the associated
transport equation $V = R_n (I-I_{c}) = R_n I_{bulk}$ implies that
$I_c$ does not participate in the dc voltage response. In other
words, the flux spots move in the surface sheath while being in
their critical state most of the time.

  This dynamic behavior corresponds to the
mean motion and explains the main dc properties. In addition to
this mean-field like picture, one can speculate that, during the
flux spots motion, many instabilities occur close the surface due
to the local release of boundary conditions \cite{placais}. The
surface current is time-dependent and can be described by its mean
value $i_c$ and its standard deviation $u^{*}$. The latter
represents the statistical fluctuation's magnitude. One expects
large fluctuations ($u^{*} \lesssim i_c$), correlated in a typical
size $c$ larger than the intervortex distance $a_0$. The scale of
the measurement is given by $S$, the surface contained between the
contacts. $N = S/c^{2}$ statistically independent fluctuators have
to be considered, and their statistical averaging reduces the
apparent fluctuations to $\delta i_c^{*} = u^{*} / \sqrt{N} $. Up
to now, only the surface current flowing in a very thin layer
(about $\xi \approx$ 30 nm compared to a sample thickness of
$10^{4} \times \xi$) is the fluctuating quantity. The measured
voltage noise is $\delta V^{*}=R_{n} \delta I_{bulk}^{*}$. Now the
current conservation implies that surface current fluctuations are
counterbalanced by bulk current fluctuations $\delta
I_{bulk}^{*}=\delta I_{c}^{*} \approx 2W \delta i_c^{*}$. Finally,
the rms noise is given by:
\begin{equation}
  \delta V^{*}=R_{n} \delta I_c^{*}=R_n I_{c} \sqrt{c^{2}/S}
\end{equation}

Experimental signatures of this mechanism would be the existence
of large voltage noise even without bulk vortices, i.e., in the
surface superconducting regime, and the verification this noise is
controlled by the surface critical current.

 One of the most critical points in this
experiment is to identify properly the surface superconducting
regime. Several experimental techniques were performed for this
purpose (fig. 1). Specific heat measurements give a robust
localization of the second critical field, i.e., the field below
which bulk superconductivity appears. Under liquid Helium
temperature (T= 4.2K), we find that the onset of the bulk
superconducting signal is at $Bc_2 = 0.295 \pm 0.005 T$  (fig. 1
top), in good agreement with value representative of pure Niobium
\cite{desorbo}. The magnetization curve shows the existence of
small hysteretic currents $i_c$ above $Bc_2$ (fig. 1 middle),
which are characteristics a surface superconducting state
\cite{fink}.  All the departure from the Ohmic normal state
behavior is due to these surface currents \cite{kim}. At the same
time, the differential resistance $R$ is $dV/dI \approx R_n$  at
high current even if the normal state is not reached, as simply
state by $V/I \neq R_n$ (fig. 1 bottom). Such a small difference
between $R$ and $R_n$ can naturally be attributed to the smallness
of the surface sheath volume, where the superconducting order
parameter can relax.
 $Bc_3$ can be
estimated from the disappearance of $i_c$, here at $0.52 \pm 0.02$
T. The obtained ratio $Bc_{3}$/$Bc_{2}$ = $1.76 \pm 0.06$ is close
to the theoretical value $1.69$. Since the critical region where
surface superconductivity takes place has been delimited, we can
investigate if an excess noise exists in this regime and
eventually analyze its behavior.

 The principle of the conduction noise measurements is the following one. The temporal evolution
of the voltage is digitally acquired after ultra low noise
differential amplification (NF-SA 400F3) with usual care to
eliminate external interference \cite{jojo} (Fig. 2). It is then
Fourier Transformed and squared in order to obtain the Power
Spectral Density of the voltage fluctuations $\tilde{V}^{2}(f)$.
The final resolution at low frequency is 0.7 nV/$\sqrt{Hz}$. Shown
in the figure 3 (Top) is the variation of $\delta V^{*} =
\sqrt{\int \tilde{V}^{2}(f) df}$, integrated over the low
frequency bandwidth (6 Hz- 200 Hz), as function of the magnetic
field. For $ B \geq B_{c3}\approx 0.52T$, no measurable noise is
observed, meaning that the metallic part (and the contacts) of the
sample can be considered as noise-free within our resolution. This
confirms that $R_n$ is not a relevant fluctuating quantity in this
experiment. For $B \lesssim B_{c3}$, the noise appears and grows
when the field is decreased. Also shown in figure 3 (Bottom) is
the variation of the surface critical current $i_{c} (A/cm)$
measured by V-I characteristics. In viewing the data, it is clear
that both $\delta V^*$ and $i_c$ have a similar field dependance.
 This suggests a
simple proportionality between $i_c$ and the amount of voltage
noise. This is expected from equation (1) in the simplest case
where $c$ is a constant, and a numerical application gives $c
\approx  0.8 \pm 0.2 \mu m$. It is important to note that a
similar analysis leads to $c \lesssim 1 \mu m$ in the bulk vortex
state of conventional type II superconductors \cite{placais,jojo}.
The voltage noise in the surface superconducting state appears  to
have the same magnitude as in the conventional mixed state. The
typical variation of $\delta V^{*}$ versus the transport current
is shown in figure 4. This closely mimics the noise variation
during the depinning transition a bulk vortex lattice \cite{clem}.
Thus, as proposed in \cite{placais,jojo}, there should be a
general explanation of the moving vortex noise in terms of pure
surface current fluctuations.

In order to confirm the role of the large surfaces for the noise
generation, we have modified the top and bottom surfaces
properties using a low energy ion irradiation ($Ar^{+}$ ions with
a kinetic energy of 600 eV and 30 minutes exposure). Because of
the short range of the ions, the damage is restricted in the first
nm  of the sub-surface region. Bulk properties ($\rho_n$,
$B_{c2}$, $T_c$) are consistently measured as unchanged. The
genuine irradiation effect on the surface superconductivity
properties can be a matter of debate. AFM and MEB inspections
confirm that the surface topography exhibits a strong increase of
its roughness at low spatial scale \cite{notebene}. The increase
of the surface roughness results in stronger flux spots pinning
ability, but at the same time the strength of the surface sheath
is reinforced because the roughening process multiplies the
possibilities that superconductivity nucleates at the scale of the
coherence length ($\xi \approx 30 nm$). Furthermore, as shown in
figure 4, the third critical field has changed and increased up to
at least $0.6 T$, likely because of the concentration of
impurities (implying a reduced electron mean free path) near the
surface. Whatever the genuine reason is, the central point is that
an impressive increase of the surface critical current is observed
(fig. 5), whereas the bulk properties are unchanged. Interestingly
for applications, one obtains large critical currents values ($i_c
\approx 10 A/cm$ or $J_c \approx 8.10^{4} A/cm2$ if expressed like
a current density at the sample scale) for fields much higher than
$Bc_{2}$. As shown in figure 6, an associated increase of the
noise is observed, giving more weight to the evidence of its pure
surface origin. Furthermore, quantitative analysis yields $c = 1
\pm 0.2 \mu m$ after the irradiation, very close to the fluctuator
size found for the virgin sample. This means that almost all the
increase of noise comes from the increase of the current
fluctuation magnitude, and that this latter is given by the
critical current value. A more detailed discussion of the effect
of surface irradiation for the pinning and vortex noise in the
surface superconducting regime and in the mixed state ($B \leq
Bc_{2}$), including the statistical study of the noise spectra,
will be discussed widely elsewhere.

To conclude, voltage noise due to flux spots motion in the
superconducting surface sheath has been observed. Current
conservation induces bulk noisy current whereas the noisy sources
were shown to be clearly localized close to the surface. Noise is
found to be of the same magnitude as in the conventional mixed
state of type II superconductors, and it behaves similarly. This
emphasizes the fundamental role of the boundaries in the
non-linear response of vortices.

 Acknowledgments:
This experiment was supported by "la r$\acute{e}$gion basse
Normandie".

%%%%%%%%%%%%%%%%%   BIBLIOGRAPHIE   %%%%%%%%%%%%%%%%%%%%%%%%%%%%

 \newpage
\begin{figure}[tbp]
\centering \includegraphics*[width=6cm]{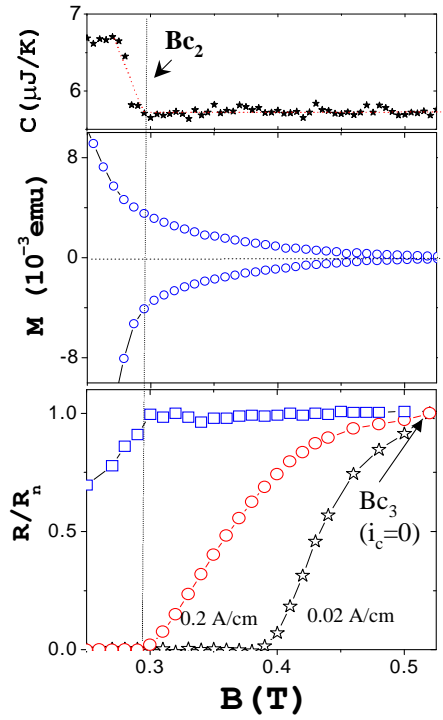}
\caption{Evidence of surface superconductivity at T= 4.2K in the
Niobium strip). Top: Specific heat as function of B. Middle:
Magnetization measured by a SQUID magnetometer. Note the
hysteresis above $Bc_{2}$. Bottom: Resistance normalized by its
normal state value. (stars: $i= 0.02 A/cm$, circles: $i= 0.2
A/cm$, squares: differential resistance for $i\gg i_c$). The
non-Ohmic behavior is due to the surface critical current.}
\label{fig.1}
\end{figure}

\begin{figure}[tbp]
\centering \includegraphics*[width=6cm]{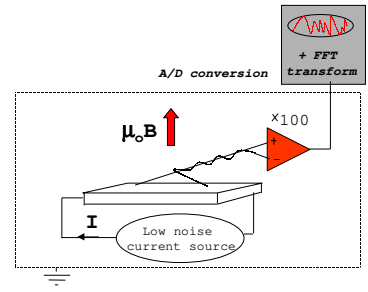}
\caption{Experimental set-up used to measure the voltage noise in
a Niobium strip (size L=5*W=1.5*t=0.2 mm$^{3}$).} \label{fig.2}
\end{figure}

\begin{figure}[tbp]
\begin{center}
\includegraphics*[angle=0,width=6cm]{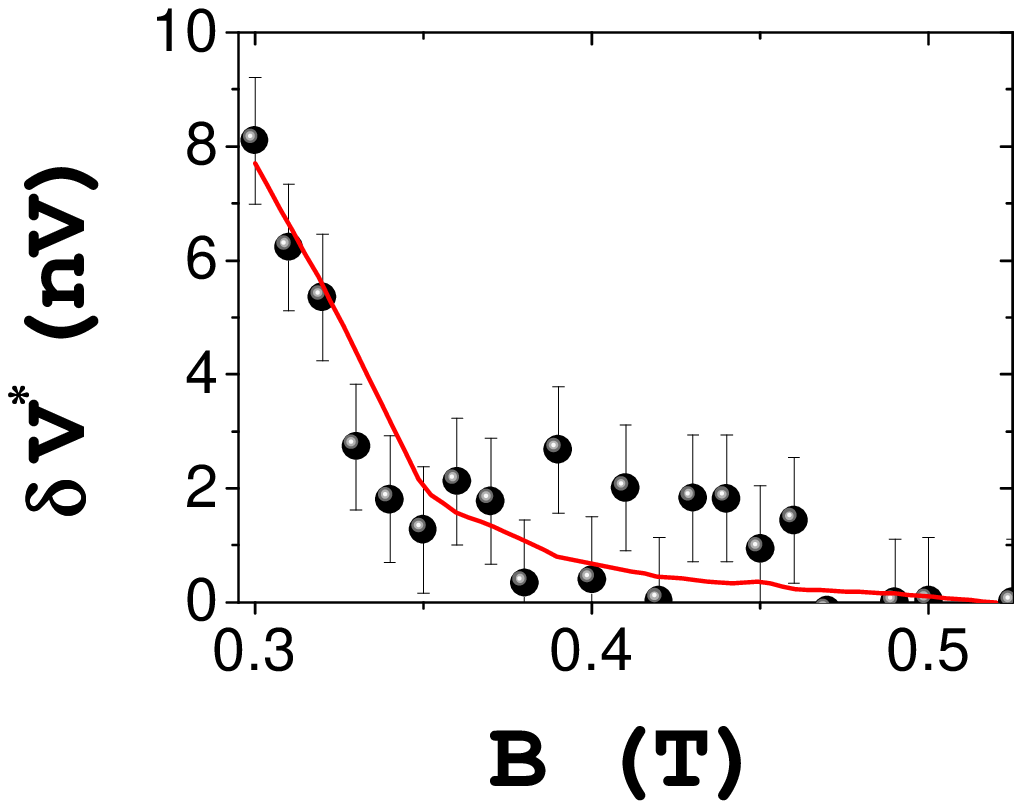}
\end{center}
\begin{center}
\includegraphics*[angle=0,width=6cm]{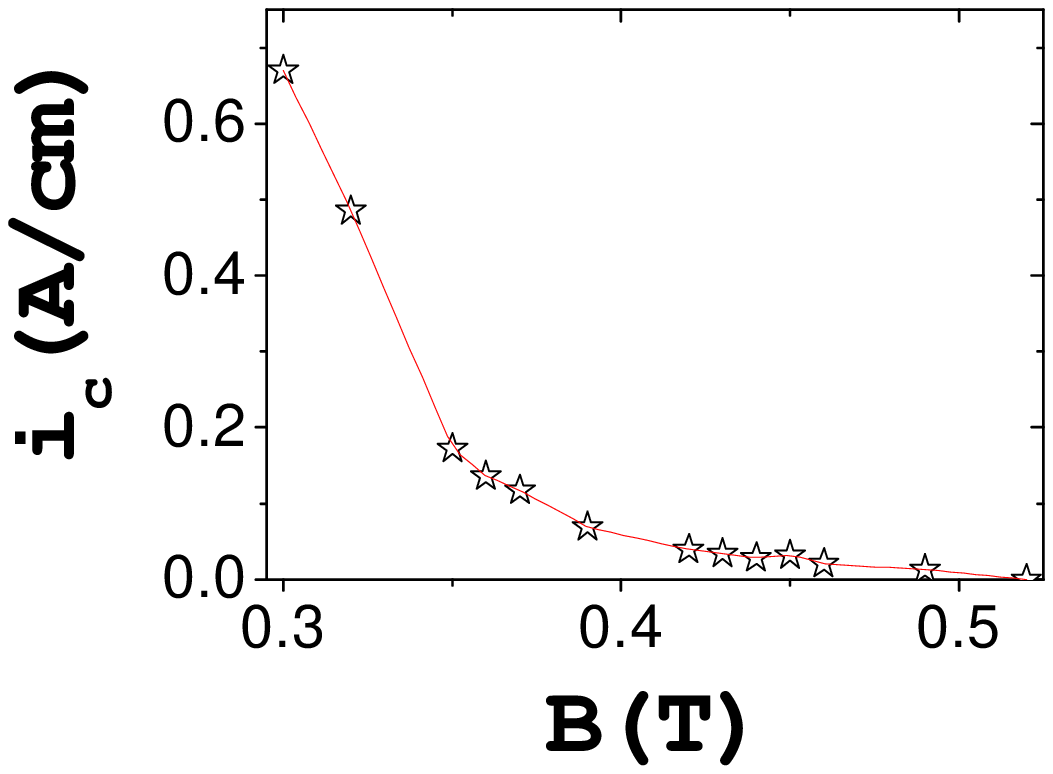}
\end{center}
\caption{Top: Variation of the voltage noise measured in the
quasi-linear regime of flux spots motion for $Bc_{2}\leq B \leq
Bc_{3}$ (T=4.2K). Also shown the master curve corresponding to the
variation of $i_c(B)$ . Bottom: Variation of the surface critical
current for $Bc_{2}\leq B \leq Bc_{3}$ (T=4.2K). The dotted line
is a guide for the eyes.}\label{fig2}
\end{figure}

\begin{figure}[tbp]
\centering \includegraphics*[angle=0,width=6cm]{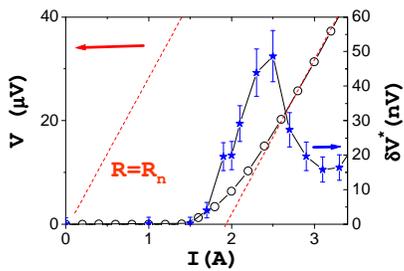}
\caption{ The voltage and the voltage noise as function of the
current (T=4.2K, B=0.36T, irradiated surfaces). Note the depinning
peak followed by the more quiet behavior characteristic of a
depinning transition. The flux-flow noise (see fig. 5) corresponds
to the regime after the depinning peak. This corresponds to the
steady state of the motion.} \label{fig3}
\end{figure}

\begin{figure}[tbp]
\begin{center}
\includegraphics*[angle=0,width=6cm]{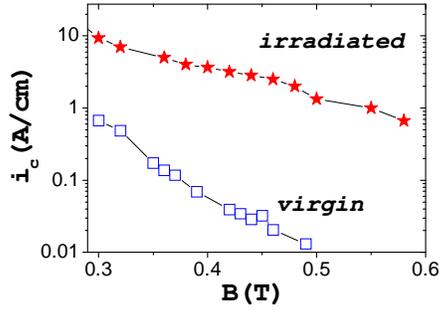}
\end{center}
\caption{Surface critical current for $Bc_{2}\leq B \leq Bc_{3}$
for the virgin and for the irradiated samples (T=4.2K). Note the
huge increase of $i_c$ after the surface irradiation.}
\label{fig.4}
\end{figure}

\begin{figure}[tbp]
\centering \includegraphics*[angle=0,width=6cm]{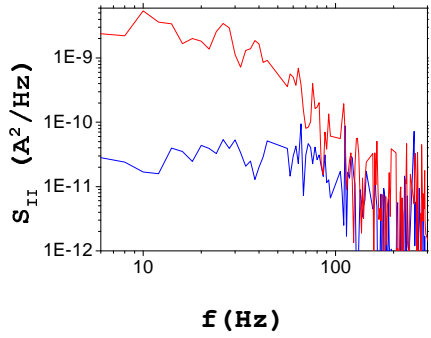}
 \caption{Typical excess of flux flow noise (the units correspond to the spectral density of the elementary
surface current fluctuations $S_{II}=S_{vv}.S/(2.W.R_n)2$)
(T=4.2K, B =0.36T). Top curve: irradiated sample. Bottom curve:
virgin sample. The integration of these spectra over the 6Hz-200Hz
bandwidth gives $i_c.c$, the value of the elementary fluctuation
(in Amperes). $c$ is deduced from this value.} \label{fig5}
\end{figure}


\begin{references}
 \label{sec:TeXbooks}

\bibitem{vanderziel} A. Van Der Ziel, fluctuations phenomena in semiconductors, Edited by C.A.
Hogarth, London, Butterworths scientific publications (1959).

\bibitem{dutta} P. Dutta and M. Horn, Rev. Mod. Phys. 53, 497 (1981).

\bibitem{zimmerman} N. M. Zimmerman, J. H. Scofield, J. V. Mantese, and W. W. Webb, Phys. Rev. B 34, 773
(1986).

\bibitem{gruner} G. Gr$\ddot{u}$ner, Rev. Mod. Phys. 60, 1129 (1988).

\bibitem{zeldov}Y. Paltiel, E. Zeldov, Y. Myasoedov, M. L. Rappaport, G. Jung, S. Bhattacharya, M. J. Higgins, Z. L. Xiao, E. Y. Andrei, P. L. Gammel, and D. J. Bishop
Phys. Rev. Lett. 85, 3712 (2000).

\bibitem{nono} N. L$\ddot{u}$tke-Entrup, B. Pla\c{c}ais, P. Mathieu, and Y. Simon
Phys. Rev. Lett. 79, 2538-2541 (1997).

\bibitem{ong} N. P. Ong, G. Verma, and K. Maki, Phys. Rev. Lett. 52, 663 (1984).

\bibitem{echo} E. M. Forgan, P. G. Kealey, S. T. Johnson, A. Pautrat, Ch. Simon, S. L. Lee, C. M. Aegerter, R. Cubitt, B. Farago, and P. Schleger
Phys. Rev. Lett. 85, 3488 (2000).

\bibitem{jos} B. D. Josephson, Phys. Lett. 16, 242 (1965).

\bibitem{clem} J. R. Clem, Phys. Rep. 75, 1 (1981).

\bibitem{paltiel} Y. Paltiel, G. Jung, Y. Myasoedov, M. L. Rappaport, E. Zeldov, S. Bhattacharya, and M. J. Higgins, Fluctuation and Noise Lett. 2, 31 (2002).

\bibitem{placais}
B. Pla\c{c}ais, P. Mathieu, and Y. Simon, Phys. Rev. Lett. 70,
1521 (1993).

\bibitem{degennes} D. Saint James and P.G. De Gennes, Phys. Lett. 7, 306 (1964).

\bibitem{desorbo} W. DeSorbo, Phys. Rev. 135, A1190 (1965).

\bibitem{kulik} I.O. Kulik, Sov. Phys. JETP 28, 461 (1969).

\bibitem{patrice} P. Mathieu, B. Pla\c{c}ais and Y. Simon, Phys. Rev. B 48, 7376 (1993).

\bibitem{hart} H.R. Hart, Jr. and P.S. Swartz, Phys. Rev. 156, 403 (1966).

\bibitem{fink} H. J. Fink and L. J. Barnes, Phys. Rev. Lett. 15, 792 (1965).

\bibitem{kim} C. F. Hempstead and Y. B. Kim, Phys. Rev. Lett. 12, 145 (1964).

\bibitem{jojo} J. Scola, A. Pautrat, C. Goupil and Ch. Simon, Phys. Rev. B 71, 104507 (2005).

\bibitem{notebene} The roughness has been controlled by AFM measurements. We have found
0.7 nm rms before the irradiation and 2.2 nm rms after the
irradiation (measurements averaged over a length of 1 $\mu$m).

\end{references}
\end{document}